\documentstyle[12pt]{article}
\begin{document}

\begin{center}
{\large INTEGRABILITY IN FLUID DYNAMICS}\\
\vskip 1cm
Subir Ghosh\\
\vskip .5cm
Physics and Applied Mathematics Unit,\\
Indian Statistical Institute,\\
203 B. T. Road, Calcutta 700035, \\
India.
\end{center}
\vskip 1cm
{\bf Abstract:}\\
 3+1-dimensional free inviscid fluid dynamics is
shown to satisfy the criteria for exact integrability, {\it i.e.}
having an infinite set of independent,
conserved quantities in involution, with the Hamiltonian being one
of them. With (density dependent) interaction present,
 distinct infinite serieses of conserved quantities
 in involution are discovered. Clebsch
  parametrization of the velocity field is used in the 
 the latter analysis. Relativistic
 generalization of the free system is also
 shown to be integrable.
----------------------------------------------------------------------
\vskip 1cm
The necessary conditions for a (classical) field theory to be
completely integrable in {\it more than one} spatial dimensions
 are
expected to correspond to their 1+1-dimensional counterpart:
existence of an
infinite
number of independent conserved quantities in involution,
(meaning that they commute
at the Poisson Bracket (PB) level) with the Hamiltonian being one
of the above \cite{das}. These features
generally induce  soliton like solutions, which are
physically interesting, both in 
(non-relativistic) condensed matter theory and in
(relativistic)  high energy physics, for simulating particle-like
excitations. However, examples of 
 integrable models in higher
dimension are quite rare \cite{fa}. In the
present Letter, we study various aspects of integrability
in the inviscid fluid model
in 3+1-dimensions, both with and without interaction \cite{sg1}.
Specifically, we show that {\bf(i)} the free  fluid models, compressible
  and incompressible, are  integrable, {\bf (ii)} compressible
  fluid model with
(density dependent) interaction  has two distinct infinite
serieses of independent conserved quantities in involution and
{\bf (iii)}  a relativistic generalization of the free fluid
model is also integrable. "Free" model means that there is no
pressure term in the equations of motion.

Exact integrability in 1+1-dimensions has been proved \cite{baz}
by exploiting the
connection between fluid models with (Chaplygin type of)
density dependent interactions and brane system \cite{j1}. Conservation
laws for irrotational fluids in higher dimensions have also been studied
\cite{baz}.

The Hamiltonian system with the basic PBs and equations of motion are
the following \cite{th}, 
\begin{equation}
 {\cal H}={1\over 2}\rho v_iv_i +R(\rho )~,~~\{v_i(x),\rho (y)\}={{\partial }
 \over {\partial x^i}}\delta (x-y)~,~~\{v_i(x),v_j(y)\}=
 -{{\partial _iv_j-\partial _jv_i}\over {\rho }}\delta (x-y).
 \label{4}
 \end{equation}
\begin{equation}
\partial _t \rho =\{\rho (x),H\} =\partial _i(\rho v_i)~,~
\partial _t v_i=\{v_i(x),H\}=
(v_j\partial _j)v_i+\partial _i({{\partial R}\over {
\partial \rho }}),
\label{1}
\end{equation}
where $\rho $ and $v_i$ denote the density and velocity fields
respectively, $i=1,2,3$ and a density dependent pressure term has been considered.
Immedietely we notice that
for the free theory, $(R=0)$, for  arbitrary smooth functions $F(v_j)$
and $G(v_j)$,
\begin{equation}
\partial _t(\rho F(v_j))+\partial _i(\rho F(v_j)v^i)=0,
\label{int}
\end{equation}
\begin{equation}
\{\int d^3x(\rho F),\int d^3y(\rho G)\}=\int d^3x\partial _j
(\rho F{{\partial G}\over {\partial v_j}}) =0.
\label{inv}
\end{equation}
(\ref{int}) constitute an infinite number of conservation laws and
(\ref{inv}) shows that the corresponding charges are in involution.
Obviously the free ${\cal H}$  belongs to this set. Hence the free theory
satisfies the criteria of exact integrability. However,
in presence of the interaction, $R\ne 0$, the conservation law (\ref{int})
does not hold \cite{6a}. To study the effects
of interaction a better framework is provided by
the Clebsch parametrization of the velocity field
\cite{th}, $
 v_i(x)\equiv\partial _i\theta (x) +\alpha (x)\partial _i\beta (x)
$,  with the non-zero PBs \cite{tu},
 \begin{equation}
\{ \theta (x),\alpha (y)\}=-{\alpha \over \rho }\delta (x-y)~,~
 \{\beta (x),\alpha (y)\}={1\over\rho }\delta (x-y)~,~
\{ \theta (x),\rho (y)\}=\delta (x-y).
 \label{6}
 \end{equation}
Time evolution of the Clebsch variables are
\begin{equation}
\dot\alpha =v_i
 \partial _i\alpha  ~;~  \dot\beta =v_i \partial _i\beta ~;~ 
 \dot\theta =v_i (\partial _i\theta -{{v_i}\over 2})
 +{{\partial R}\over {\partial \rho }} ~;~
\dot\rho = \partial _i (\rho v_i).
\label{a}
\end{equation}
Using above relations, one can easily check that
\begin{equation}
\partial _t (A(\alpha )\rho )=\partial _i(A(\alpha )\rho v_i )~;~
\partial _t (B(\beta )\rho )= \partial _i(B(\beta )\rho v_i )~;~
\partial _t (C(\alpha ,\beta )\rho )= \partial _i(C(\alpha ,\beta )
\rho v_i ),
\label{d}
\end{equation}
where $A$, $B$ and $C$ are smooth functions.
The above constitute the three independent serieses of local
current conservation equations. Clearly the conserved charges
$A[\alpha ]=\int (A(\alpha )\rho )$ etc. are trivially involutive,
\begin{equation}
\{A_1[\alpha ],A_2[\alpha ]\}=\{B_1[\beta ],B_2[\beta ]\}=0,
\label{f}
\end{equation}
but the rest of the PBs between charges are non-vanishing. Hence one
can take either the set $A[\alpha ]$ or the set
$B[\beta ]$ as the independent
and conserved
charges in involution. This constitues analysis of the interacting case.
Notice that ${\cal H}$ does not belong to any of the three sets as it
contains $\theta $. It would be interesting if a bi-Hamiltonian
structure \cite{nut} could be found, in which a term from
one of the above serieses will act as Hamiltonian to reproduce
(\ref{a}) via a PB algebra other than (\ref{6}).

In this connection, the incompressible fluid turns out to be
integrable. We introduce the constraint $\rho (x)-\rho _0\approx 0$
which generates another constraint $\partial _iv_i\approx 0$
due to the equations of motion (\ref{1}). 
One can replace $\theta =-(\partial _i\partial _i)^{-1}\partial _j
(\alpha \partial _j \beta )$ from ${\cal H}$ so that it belongs
to $C(\alpha ,\beta )$. Explicitly performing the Dirac analysis
\cite{di} of the above second class pair of constraints \cite{tu1}
one can check that the conservation laws in (\ref{d})
and the corresponding charges become
$\rho $-independent.
Hence
the Hamiltonian and either the set $A(\alpha )$ or the set $B(\beta )$
constitute the conserved quantities in involution.

Some peculiarities are involved in the Noether description of the
above conserved currents (\ref{d}) from the Lagrangian  \cite{j1}
\begin{equation}
{\cal L}=
 \dot \theta \rho +\dot \beta
\alpha \rho - {\cal H},
\label{g}
\end{equation}
which correctly reproduces the PBs  (\ref{6}) by Dirac constraint analysis
 \cite{di} or symplectic
procedure \cite {fj}.
Possibly due to the non-canonical and first
order nature of the system, the action respects the symmetry
transformations leading to the currents in a restricted way, where
the arbitrary functions $A(\alpha )$, $B(\beta )$ and
$C(\alpha ,\beta )$ have to be replaced by polynomials $\alpha ^n$ etc
\cite{sg1}.

In the relativistic generalization,
the free Lagrangian is expressed as \cite{j2}
\begin{equation}
{\cal L}_{rel}=j^\mu a_\mu -(j^\mu j_\mu )^{{1\over 2}}~;~
a_\mu =\partial _\mu \theta +\alpha \partial _\mu \beta ~,~
j^\mu =(\rho ,~\rho v^i) .
\label{k}
\end{equation}
Expanding  the square root as 
$\rho (1+v^iv_i)^{{1\over 2}}\approx \rho (1+{1\over 2}v^iv_i +...),$
and dropping the uninteresting $\int \rho $ term
\cite{j2}, (since it can only
influence the time evolution of $\theta $ by a constant translation),
 we can recover the non-relativistic Lagrangian. We emphasise that
the symplectic structure and the associated PB algebra remains
 unaltered from (\ref{6}). The   Hamiltonian now is modified to
\begin{equation}
{\cal H}_{rel}=\rho [v_iv_i+(1-v_iv_i)^{{1\over 2}}],
\label{l}
\end{equation}
which changes the equations of motion to the following:
$$\dot\alpha =L_i\partial _i\alpha ~;~\dot\beta =L_i\partial _i\beta ~;~
\dot\theta =-L_i\alpha \partial _i\beta + 
[v_iv_i+(1-v_iv_i)^{{1\over 2}}]
~;~\dot\rho=\partial _i(\rho L_i),$$
\begin{equation}
L_i={1\over \rho }{{\partial {\cal H}_{rel}}\over {\partial v_i}}
=v_i[2-(1-v_iv_i)^{-{1\over 2}}].
\label{m}
\end{equation}
Notice that in the lowest order, $L_i\approx v_i+O(v^3)$, the previous
equations are recovered. Clearly the previous conservation laws will 
remain intact by replacing $v_i$ by $L_i$ in the expressions
(\ref{d}) for the currents. Integrability is proved since in analogy
to (\ref{int}), we have
$\partial _t(\rho F(v_j))+\partial _i(\rho F(v_j)L^i)=0
$, (\ref{inv}) remains unchanged and ${\cal H}_{rel}$
belongs to $F(v_i)$.

Regarding the possibility of soliton solutions in the fluid model, a
comment is in order. The Virial theorem \cite{rr} proving 
absence of solitons in models in higher dimensions, involving {\it only}
scalar fields, depends on the Lagrangian having the canonical
relativistic form ${\cal L}=\partial _\mu \phi _i\partial ^\mu \phi _i
-U(\phi _j)$. The relativistic Lagrangian
(\ref{k}) and its non-relativistic limit (\ref{g}) both are
 first order in nature with
a non-canonical PB algebra between the basic scalar fields $\rho $,
 $\theta $, $\alpha $ and $\beta $. Hence, {\it a priori } it is not
obvious whether the Virial theorem is applicable here. A natural
starting point is provided in the Batalin-Tyutin \cite{bt} extension
of this model (see Ghosh in \cite{tu}), where the PBs are
canonical. All the above results are derivable here in a modified form.
Also this version is suitable for quantiztion.

 To conclude, we have shown that the 3+1-dimensional
 free fluid model satisfies the exact integrability criteria. For a
 density dependent interaction, there exist three distinct
infinite sets of
 independent, conserved charges of which terms in two sets are in
 involution. Subtleties in the Noether
 description of the conservation laws are pointed out.
  We have also shown that
 these conservation laws survive, albeit with modifications, in a
 relativistic generalization of the free theory.
   Establisment of a bi-Hamiltonian
 structure in case of interactions
 and soliton solutions in the model will be worth studying. How far
 these conclusions can be generalized to d-brane systems remains an open
 question. From another point of view, stability analysis of
 non-linear systems {\cite{mar} in Clebsch representation will
 be interesting since, (as stated in \cite{mar}), 
 a large amount of arbitrariness in the
 structure of the conserved quantities,
 which is present here, is required in the analysis.

 \vskip .5cm
 It is a pleasure to thank Dr. B. Basu Mallick
 for discussions. Also I thank Dr. D. Bazeia for correspondence.


\begin{thebibliography}{99}
\bibitem{6a} There is a conservation law for $F(v_j)\approx v_j$
 and $R(\rho )\approx \rho ^n$.
\end{thebibliography}
\end{document}